%% file: main.tex
\documentclass[sigconf,screen]{acmart}
\usepackage{verbatim}
\usepackage{caption}
\usepackage{subcaption}
\usepackage[]{algorithm2e}
\usepackage{adjustbox}
\usepackage{graphicx}
\usepackage{booktabs}
\usepackage{relsize}
\usepackage{colortbl}
\usepackage{enumerate}
\usepackage{amsmath}
\usepackage{tikz}
\usepackage{pgf}
\usepackage{pbox}
\usepackage{rotating}
\usepackage{multirow}
\usepackage{balance}
\usepackage{tablefootnote}
\usepackage{flushend}
\usepackage{float}
\usepackage{threeparttable}
\usepackage{tcolorbox}
\usepackage{stackengine,scalerel}
\usepackage{cleveref}[2012/02/15]
\usepackage{arydshln}
\usepackage{microtype}

\crefformat{footnote}{#2\footnotemark[#1]#3}

\usepackage{soul}

\usepackage{listings}

\usepackage{xcolor}
\usepackage{cuted}

\AtBeginDocument{%
  \providecommand\BibTeX{{%
    \normalfont B\kern-0.5em{\scshape i\kern-0.25em b}\kern-0.8em\TeX}}}

\newcommand{\startlist}{\begin{list}{\labelitemi}{\leftmargin=1em}\setlength{\itemsep}{-1mm}}
\newcommand{\stoplist}{\end{list}}

\usepackage{tikz}

\newcommand{\resultBox}[1]{\begin{center}\noindent\fbox{\parbox{0.95\linewidth}{\textit{#1}}}\end{center} }
\newcommand{\smallsection}[1]{\noindent {\bf{#1}}.\hspace{1mm}}
\newcommand{\ea}{{\em et al.}}
\newcommand{\com}{Atlassian}

\usepackage{tcolorbox}

\newcommand\dddagger{%
  \sbox0{\ddag}\scalerel*{%
  \stackengine{-.6\ht0}{\ddag}{\ddag}{O}{c}{F}{F}{S}}{\ddag}%
}

\newcommand{\rqone}{What are the most important factors associated with CI Build Failures at Atlassian?}
\newcommand{\rqtwo}{What are practitioners’ perceptions about CI build failures?}
\newcommand{\rqthree}{What are practitioners' perceptions on the usefulness of CI build prediction?}
\newcommand{\rqfour}{What are practitioners’ perceptions on the explanations and suggestions of CI build prediction?}



\newcommand{\approach}[1]{\textsc{RevSpot}}

\setcopyright{acmlicensed}
\acmDOI{10.1145/3663529.3663856}
\acmYear{2024}
\copyrightyear{2024}
\acmSubmissionID{fsecomp24industry-p93-p}
\acmISBN{979-8-4007-0658-5/24/07}
\acmConference[FSE Companion '24]{Companion Proceedings of the 32nd ACM International Conference on the Foundations of Software Engineering}{July 15--19, 2024}{Porto de Galinhas, Brazil}
\acmBooktitle{Companion Proceedings of the 32nd ACM International Conference on the Foundations of Software Engineering (FSE Companion '24), July 15--19, 2024, Porto de Galinhas, Brazil}
\received{2024-02-08}
\received[accepted]{2024-04-18}

\begin{document}
\graphicspath{{figures/}}

\title{Practitioners' Challenges and Perceptions of CI Build Failure Predictions at Atlassian} 




\author{Yang Hong}
\email{yang.hong1@monash.edu}
\affiliation{
    \institution{Monash University}
  \country{Australia}
}
\author{Chakkrit Tantithamthavorn}
\email{chakkrit@monash.edu}
\affiliation{
    \institution{Monash University}
  \country{Australia}
}
\authornote{The corresponding author}

\author{Jirat Pasuksmit}
\email{jpasuksmit@atlassian.com}
\affiliation{
    \institution{Atlassian}
  \country{Australia}
}

\author{Patanamon Thongtanunam}
\email{patanamon.t@unimelb.edu.au}
\affiliation{
    \institution{The University of Melbourne}
  \country{Australia}
}

\author{Arik Friedman}
\email{afriedman@atlassian.com}
\affiliation{
    \institution{Atlassian}
  \country{Australia}
}

\author{Xing Zhao}
\email{xzhao2@atlassian.com}
\affiliation{
    \institution{Atlassian}
  \country{Australia}
}

\author{Anton Krasikov}
\email{akrasikov@atlassian.com}
\affiliation{
    \institution{Atlassian}
  \country{Australia}
}

\renewcommand{\shortauthors}{Hong et al.}

\sloppy

\begin{abstract}
Continuous Integration (CI) build failures could significantly impact the software development process and teams, such as delaying the release of new features and reducing developers' productivity.
In this work, we report on an empirical study that investigates CI build failures throughout product development at Atlassian.
Our quantitative analysis found that the repository dimension is the key factor influencing CI build failures.
In addition, our qualitative survey revealed that Atlassian developers perceive CI build failures as challenging issues in practice.
Furthermore, we found that the CI build prediction can not only provide proactive insight into CI build failures but also facilitate the team's decision-making. 
Our study sheds light on the challenges and expectations involved in integrating CI build prediction tools into the Bitbucket environment, providing valuable insights for enhancing CI processes.

\end{abstract}


   

\maketitle
\begingroup\renewcommand\thefootnote{\textsection}
\endgroup

\input{sections/introduction}
\input{sections/background}
\input{sections/approach}
\input{sections/result}

\input{sections/discussion}

\input{sections/threat}
\input{sections/related}
\input{sections/conclusion}

\bibliographystyle{ACM-Reference-Format}
\bibliography{references}

\end{document}

%% file: sections/introduction.tex
\vspace{-3mm}

\section{Introduction}

Continuous integration (CI) is a software development practice where developers frequently merge their code into a shared repository~\cite{schneider2008continuous,beller2017travistorrent}. 
Each integration undergoes a verification process through an automated build, including code dependency installation, code compilation, and test case execution. 
The benefits of CI for software organizations are substantial; e.g., it enables earlier and faster identification and resolution of integration errors~\cite{schneider2008continuous}, enhances developer productivity~\cite{hilton2016usage}, improves product quality~\cite{vasilescu2015quality} and reduces development and delivery time~\cite{zhao2017impact}.

Despite these advantages, CI comes with its own challenges.
At Atlassian, one of the challenges is to reduce CI build failures on the main branch.
CI build failures on the main branch could significantly impact the software development process and development teams.
For example, they could delay rollouts of new features or security patches, require rollbacks to fix the issues, and hamper the productivity of software engineers~\cite{ananthanarayanan2019keeping}. 
From 2021 to 2023, we observed that failed CI builds on the main branch resulted in an average of 120 hours of wasted build time per project per year in the studied Atlassian's projects.
Such disruptions not only prevent developers from proceeding further with the development but also incur a high cost for organizations to rectify.
Therefore, this challenge necessitates a thorough investigation of the CI build failures at Atlassian.

To mitigate CI build failures, various CI build prediction techniques~\cite{chen2020buildfast,saidani2020predicting,santos2022investigating,kawalerowicz2023continuous,saidani2022improving,xia2017could} have been developed to preemptively detect states in the CI process that are likely to cause build failures. This enables developers to take the necessary actions and avoid these failures.
However, despite the availability of these techniques, there still remain gaps between industry and research.
First, prior studies are based on open-source projects, which may not generalize to the industrial context with different scales.
In addition, while explainability is an important aspect of their adoption by developers~\cite{tantithamthavorn2021explainable,jiarpakdee2020empirical,jiarpakdee2021practitioners,tantithamthavorn2015impact, pornprasit2021pyexplainer, tantithamthavorn2020explainable}, little is known how developers perceive about the CI build failure predictions and their explanations.

In our study, we conducted empirical studies to investigate the factors that are associated with CI build failures and practitioners' perceptions of utilizing CI build predictions at Atlassian.
Specifically, we analyzed 350,037 pull requests from 1,630 projects at Atlassian and investigated 11 factors that are possibly associated with CI build failures.
We then conducted a survey study to gain a better understanding of the adoption of CI build failure predictions. The survey had 53 respondents (i.e., Atlassian practitioners who participated).
We answered four research questions below:

\begin{enumerate}[{\bf RQ1)}]
\item {\bf \rqone}\\
We found that factors in the repository dimension (e.g., the ratio of the builds that failed out of the recent five
builds of the repository and the ratio of all historical builds of the repository that failed) play the most important role in influencing CI build outcomes.
\item[{\bf RQ2)}] {\bf \rqtwo}\\
From the respondents' perspective,  while CI build failures could impact the development process and productivity of the team, they could also provide the opportunity for developers to learn from build failures and avoid more serious issues in the future.

\item[{\bf RQ3)}] {\bf \rqthree}\\
Respondents perceived that CI build predictions could provide proactive insight and aid in the team's decision-making.
However, they noted several challenges that need to be addressed before adopting the predictions in practice, such as the potential for developers' over-reliance on them.

\item[{\bf RQ4)}] {\bf \rqfour}\\
Respondents agreed with explanations and suggestions based on the percentage of current changed files that were involved in the previous failed builds and the number of changed files.
Developers pointed out that the explanations and suggestions for the related factors could help developers understand CI build failure.
However, they also mentioned that the explanations and suggestions should be optimized to fit the context of development scenarios.
\end{enumerate}

Our study sheds light on the challenges and expectations involved in integrating CI build prediction tools, providing valuable insights for enhancing CI processes.





%% file: sections/background.tex
\section{The CI build process at Atlassian} \label{sec:background}

At Atlassian, developers use Bitbucket\footnote{https://bitbucket.org/product} for code review and the Continuous Integration (CI) workflow.
Initially, a developer submits code changes through a pull request (PR), followed by a code review process to check the quality of code changes. 
During this process, the developer scrutinizes and improves the code based on the feedback from reviewers. 
Once reviewers approve the PR, it will proceed to the CI build process.

\begin{figure}[ht]
    \centering
    \includegraphics[width=\columnwidth]{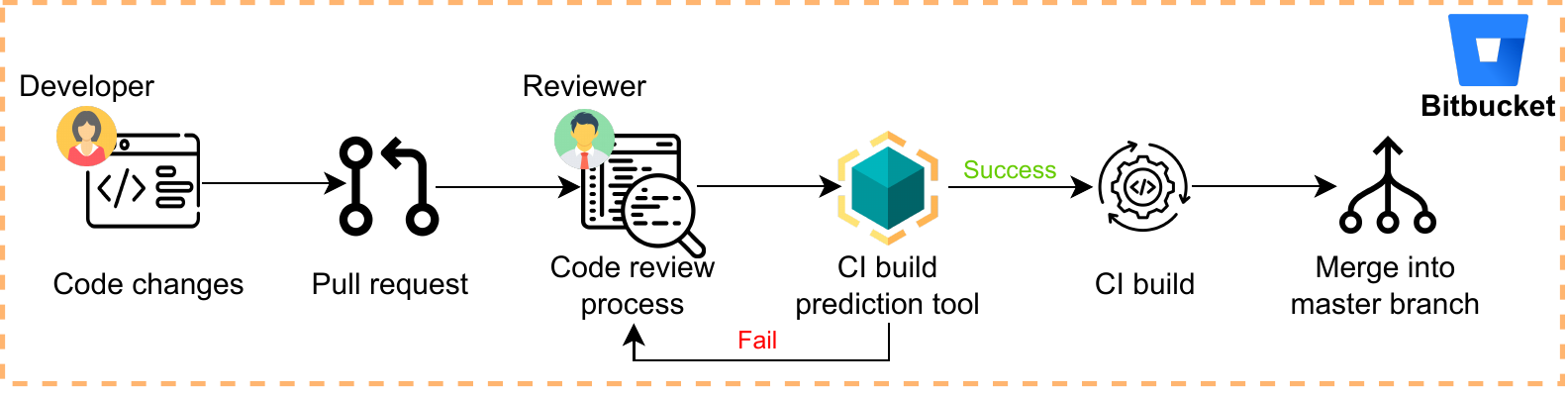}
    \caption{The potential usage scenario of the CI build predictions within the CI build process at Atlassian.}
    \label{fig:background}
\end{figure}

However, the CI build of the PR may fail even if it is carefully reviewed.
The build failure could significantly impact the software development process at Atlassian.
To mitigate such a problem, prior studies proposed various CI build prediction techniques to early predict the build outcomes (see Figure~\ref{fig:background}).
If the predictions can early indicate that the build will fail, developers can examine the PR again before sending it to the CI process.
This enables developers to take the necessary actions to avoid potential CI build failures.
If the predictions indicate that the build will succeed, the PR will be delivered to the CI build process and merged into the main branch of the code repository.

In this work, the close integration of Atlassian's development teams with Bitbucket gives the opportunity to investigate the CI build processes \textit{in situ} and gain insights into developers' perceptions and expectations if integrating CI build predictions into Bitbucket.

%% file: sections/approach.tex
\section{Research Methodology} \label{sec:approach}
In this section, we present the research methodology for our empirical study.

\subsection{Research Questions}

We present the motivation for our research questions below:
\begin{enumerate}[{\bf RQ1)}]
\item {\bf \rqone}\\
While previous studies looked into the effect of a variety of factors that influence CI build failures in open source projects~\cite{chen2020buildfast,saidani2020predicting,santos2022investigating}, there is a lack of understanding of factors that impact CI build failures in industry contexts.
Our study aims to bridge this gap by conducting a quantitative analysis of factors affecting CI build failures at Atlassian.
\item[{\bf RQ2)}] {\bf \rqtwo}\\
The CI build process is typically time and resource-consuming, as running failed builds can take hours until discovering the breakage.
Prior studies~\cite{saidani2022improving, chen2020buildfast,hilton2016usage, hilton2017trade} found that CI build failures may cause disruptions in the development process and delays in the product release dates in the context of open-source projects.
This RQ seeks to understand the impact of CI build failures at Atlassian from the developers' perspective through a survey.

\item[{\bf RQ3)}] {\bf \rqthree}\\
To adopt the CI build prediction in practice, it is necessary to understand developers' attitudes towards integrating such technique into their routine workflows.
This RQ focuses on understanding developers' perceptions of the utility of the CI build prediction in practice.

\item[{\bf RQ4)}] {\bf \rqfour}\\
The effectiveness of CI build prediction hinges on their ability to provide insightful explanations and suggestions~\cite{tantithamthavorn2021explainable,jiarpakdee2020empirical,jiarpakdee2021practitioners}.
Both explanations and suggestions are vital in helping developers understand and mitigate factors leading to CI build failures. 
We set out this RQ to understand the developers' perception of factors influencing the CI build outcome, along with their views on the effectiveness of the explanations and suggestions.
\end{enumerate}

\subsection{Studied Dataset}
To address our research questions, we conducted an empirical study on Atlassian's internal software projects. These projects are created by Atlassian development teams (not Atlassian's customers).
In these projects, Atlassian developers employ a workflow that streamlines the integration of new code changes (see Section 2). 
The data relating to this workflow is collected by Atlassian teams and stored in a database for in-depth data analysis.


To perform our empirical study, we prepare our data from \com{}'s database, as described below. 

\smallsection{Selecting Data}
To identify the studied projects, we used two following criteria:
\begin{itemize}
    \item \textbf{Criterion 1: Perform a Pull Request Process.} 
    Since the pull request process plays a vital part in the CI build process, we select the projects that apply pull requests.
    \item \textbf{Criterion 2: Conduct a CI Build Process.}
    Since the paper aims to investigate the CI build, we focus on the projects that conduct the CI build process.
\end{itemize}
For Criterion 1, we identify the usage of the pull request process in projects by checking if a project includes at least one PR.
For Criterion 2, we filter out projects that do not incorporate CI builds to integrate code changes, and include only projects that have at least one CI build record.

\smallsection{Data Cleaning}
After selecting the projects, we cleaned the data to ensure the accuracy of the study results.
We first excluded the builds that are not triggered on the main branch.
Then, we performed the following steps for data cleaning.

\begin{itemize}
    \item \textbf{Excluding incomplete PRs.} 
    We exclude the incomplete PRs (i.e., marked as \textit{open}) since the PRs may still be ongoing and it is likely that they will keep changing after the data collection time.
    Hence, we only consider \textit{closed} PRs. 
    \item \textbf{Excluding incomplete builds.} 
    We consider the builds to be incomplete based on two criteria. 
    First, we consider builds without CI build completion time as incomplete builds and exclude them from this study.
    Second, there are four types of CI build status in the studied dataset: passed, failed, errored, and canceled. 
    A canceled build status denotes that the build process was interrupted. 
    We exclude canceled builds as they are also incomplete builds.
    Then, similarly to a previous study~\cite{hassan2017change}, we consider builds with passed build status as successful builds and builds with failed or errored build status as failed builds.

\end{itemize}

\smallsection{Dataset} We collected the data of PRs merged into the main branch between January 1, 2021, and September 1, 2023.
After applying the aforementioned inclusion/exclusion criteria to the data, our dataset consists of a total of 350,037 PRs from 1,630 Atlassian projects.

\begin{table}[]
\centering
\caption{An overview of studied factors.}
\label{tab:metrics}
\begin{adjustbox}{max width=\columnwidth}
\begin{tabular}{l p{6.5cm}}
\hline
\textbf{Factor} & \textbf{Description} \\ \hline
\multicolumn{2}{l}{\textit{\textbf{PR Dimension}}} \\ \hline
changes\_num & The number of changed files in the PR. \\
comment\_num & The number of review comments during the code review process of the PR. \\
reviewer\_num & The number of reviewers who have participated in the code review process of the PR. \\
wait\_time\_to\_review & The duration in minutes from when the PR was opened to when it was reviewed. \\
reviewing\_time & The duration in minutes from the opening of the PR to its closure. \\
per\_failed\_file & The ratio of changed files in the PR that were involved in the past failed builds. \\ \hline
\multicolumn{2}{l}{\textit{\textbf{Repository Dimension}}} \\ \hline
repo\_prev\_build & The build status of the previous build of the repository. \\
repo\_rec\_build & The ratio of the builds that failed out of the recent five builds of the repository. \\
repo\_hist\_build & The ratio of all historical builds of the repository that failed. \\ \hline
\multicolumn{2}{l}{\textit{\textbf{Contributor Dimension}}} \\ \hline
team\_num & The number of teams that made contributions to the PR. \\
team\_member\_num & The number of members in the PR author's team. \\ \hline
\end{tabular}
\end{adjustbox}
\end{table}

\begin{table*}[t]
\caption{Survey questions (excluding demographics questions).}
\label{tab:survey_questions}
\begin{adjustbox}{max width=0.8\textwidth}
\begin{threeparttable}
\begin{tabular}{l|l}
\hline
Item  & Question                                                                                                                           \\ \hline
      & \textbf{The CI build failure (RQ2)}                                                                                                \\
Q1$\dagger$ &
  \begin{tabular}[c]{@{}l@{}}From your past experience, how challenging is it to resolve the CI build failures?\end{tabular} \\
Q2*   & \begin{tabular}[c]{@{}l@{}}How does CI build failures impact yourself and the organization?\end{tabular}                        \\ 
Q3$\diamond$  & \begin{tabular}[c]{@{}l@{}}In your own experience, why do CI builds fail when merging a PR into the main branch?\end{tabular} \\ \hline
      & \textbf{The CI build prediction (RQ3)}                                                                                        \\
Q4*$\ddagger$ & Do you think the CI build prediction is useful?                                                                               \\ \hline
      & \textbf{Developers' perception about factors (RQ4)}                                                                                                        \\
Q5-Q7*$\dddagger$ &
  \begin{tabular}[c]{@{}l@{}}Assuming that a CI build prediction predicts a given build with a 90\% risk of CI build failure, with \\ an explanation and a suggestion displayed in the UI. Do you agree with its explanation and suggestion?\end{tabular} \\ \hline
\end{tabular}
\begin{tablenotes}
\footnotesize
\item Questions with asterisk* included an optional open-ended answer for response or further elaboration.
\item $\diamond$: Multiple-selection question: $\square$ Code issues, $\square$ Failed tests, $\square$ Merge conflicts, $\square$ Configuration errors, $\square$ Dependency issues, $\square$ Other (Open-end option).
\item $\dagger$: Likert scale of challenging: $\bigcirc$ Extremely challenging, $\bigcirc$ Very challenging, $\bigcirc$ Moderate, $\bigcirc$ Not very challenging, $\bigcirc$ Extremely not challenging.
\item $\ddagger$: Likert scale of usefulness: $\bigcirc$ Extremely useful, $\bigcirc$ Somewhat useful, $\bigcirc$ Neutral, $\bigcirc$ Somewhat not useful, \\ $\bigcirc$ Extremely not useful.
\item $\dddagger$: Likert scale of agreement: $\bigcirc$ Strongly agree, $\bigcirc$ Agree, $\bigcirc$ Neutral, $\bigcirc$ Disagree, $\bigcirc$ Strongly disagree.
\end{tablenotes}
\end{threeparttable}
\end{adjustbox}
\end{table*}

\subsection{Explanatory Factors}
Previous works suggest a set of different factors affecting CI build outcomes~\cite{xia2017could, saidani2020predicting, ruangwan2019impact, chen2020buildfast}.
Motivated by these works, we derive explanatory factors from our dataset.
Table~\ref{tab:metrics} shows the factors we consider in our study and their corresponding descriptions. 
We categorize our factors into three dimensions: (1) PR dimension, focusing on the attributes of the PR, (2) Repository dimension, capturing information from previous builds in the repository, and (3) Contributor dimension, measuring team engagement.

Note that the scope of data available for our study is constrained. 
First, the data only contains PR level, repository level, and team-related information. 
Secondly, to strictly comply with \com{}'s privacy policies, we are required to exclude any data, including but not limited to personally identifiable data and user-generated content to ensure the protection of sensitive information. 



\subsection{Data Analysis}
Similarly to past studies~\cite{thongtanunam2017review,tantithamthavorn2018experience}, we aim to understand the effect of the selected factors on the CI build outcome of a PR at Atlassian. To this end, we fit a Logistic Regression Model (LRM). 
LRM models are commonly used to capture the relationship between the explanatory variables (i.e., factors described in Table~\ref{tab:metrics}) and a response variable (i.e., the CI build outcome). 
In our study, the response variable is assigned the value of TRUE if a build fails, and FALSE otherwise.

\smallsection{Correlation \& Redundancy Analysis}
Highly correlated explanatory variables can interfere with the results of model analysis. 
We start by employing Spearman rank correlation tests ($|\rho|$) to assess the correlation among these variables.
As suggested by Hinkle~\ea~\cite{hinkle1998analysis}, a value of Spearman correlation coefficient greater than 0.7 is indicative of strong correlations. 
Therefore, we set this value as our threshold for identifying highly correlated explanatory variables. 
Then, we use a variable clustering analysis technique~\cite{sarle1990varclus} to construct a hierarchical overview of the correlation. 
Within each cluster where $|\rho| >$ 0.7, we select one variable from the cluster for building the model.

Additionally, explanatory variables are not highly correlated but may still be redundant. 
Redundant variables in an explanatory model can distort the modeled relationship between the explanatory and response variables.
To address this, we use the \textit{redun} function from the \textit{rms} R package to detect redundant variables.
We remove the variables where the models fit with an $R^{2} >$ 0.9.

\smallsection{Model Evaluation}
To ensure our model can provide meaningful and reliable results, we use the Area Under the Receiver Operating Characteristic Curve (AUC) to assess its discriminative ability on the response variable. 
An AUC value of 1 indicates strong discriminative ability, i.e., perfect separation of the builds that fail and those that do not, while an AUC value of 0.5 indicates that the model does not discriminate better than random guessing.
We perform the optimism of the AUC using a bootstrap-derived approach~\cite{efron1986biased} to evaluate the reliability of the model.
The smaller the average optimism is, the more reliable the estimates of the fit are.
We also conduct 10-fold cross-validation experiments to mitigate the risk of overfitting and ensure the model's generalizability.

\smallsection{The Power of Explanatory Variables Estimation}
Similar to prior works~\cite{mcintosh2016empirical, thongtanunam2017review}, we employ Wald statistics to estimate the influence of each explanatory variable on the model.
We use the \textit{anova} function in the \textit{rms} R package to estimate the relative contribution (Wald $\chi^2$) and the statistical significance ($p$-value) of each explanatory variable in the model. A higher Wald $\chi^2$ value denotes a greater explanatory power of a particular explanatory variable that contributes to the estimations of the model.

\smallsection{The Analysis of Variable Influence on the Response Variable}
To understand the relationships between explanatory variables and the response variable, we use the odds ratio to estimate the partial effect that explanatory variables have on the response variable. 
The odds ratio indicates the likelihood of a build failing when the value of an explanatory variable increases. 
Specifically, a larger odds ratio suggests a larger partial effect of the explanatory variable on the likelihood of build failures. 
We analyze the relative percentage that the odds have changed corresponding to the changed value of the explanatory variable while other explanatory variables are held constant.
A positive partial effect suggests a positive relationship between the explanatory variable and the response variable (i.e., increases in the variable are associated with higher odds of CI build failure).
Conversely, a negative partial effect indicates a negative relationship (i.e., increases in the variable are associated with lower odds of CI build failure). 
The magnitude of a partial effect indicates the degree to which the odds value in our model changes in response to a variation in the explanatory variable.

\subsection{Survey Design and Participants}

Our online survey is tailored for developers engaged in CI-based software development. 
The survey aims to understand Atlassian developers' work practices and their perceptions of the CI build process.
The survey begins with demographic questions aimed at collecting general background information on participants and their experience with the CI build process.
We ask the participants about their roles, work experience, the frequency with which they submit pull requests, and how often they experience CI build failures.
Excluding demographic questions, the survey consists of 7 questions: Q1-Q3 investigate the impact and developers' perceptions of CI build failure; Q4 focuses on the developer's perception of CI build prediction; Q5-Q7 aims to understand the perception of developers of factors influencing the outcome of the CI build failure (See Table~\ref{tab:survey_questions}).
The further details of Q5-Q7 are provided in Section~\ref{sec:result}.
The survey is designed to be completed within 10-15 minutes.

Acknowledging the ethical considerations highlighted by Baltes et al.~\cite{baltes2016worse} regarding unsolicited emails to developers, we opt for a more respectful approach to participant recruitment by sending invitations to take our online survey through internal developer discussion channels. 
The survey was open for one month, from November 10 to December 8, 2023, and received 53 responses from developers experienced in the CI build process.

\smallsection{Respondent Demographics}
The respondents are software engineers (91\%) and software engineering managers (7\%).
For the experience in the IT industry of respondents , 33\% of participants had more than five years of experience, 59\% of respondents had one to five years, and 7\% of respondents had less than one year.
In terms of the frequency of submitting PRs, 46\% of respondents have submitted PR more than once a day; 41\% of the respondents submitted PR once a day.
The remaining respondents submitted PR once a week.
Regarding the frequency of respondents experiencing CI build failures, 4\% of respondents experienced failures in more than 75\% of PRs, 43\% in 50\%-75\% of PRs, 37\% in 25\%-50\% of PRs, and 16\% of respondents experienced failures in less than 25\% of PRs.

\subsection{Card Sorting}
For the analysis of open-ended survey responses, we employ a thematic analysis approach to ensure our analysis remains unbiased and free from preconceptions about the responses.
Therefore, similar to previous works~\cite{kononenko2018studying,kononenko2016code}, we used an open coding approach to identify themes and group the data.
The first author split the 53 survey responses into isolated quotes
(cards).
Each card represents a single statement distinct from other statements in a particular response to a question. 
After that, two of the other authors independently served as coders to manually label cards.
They discuss any uncertainties in the cards to share a common understanding of the data. 
Finally, similar cards are merged into common themes for further analysis.

%% file: sections/result.tex
\section{Results}\label{sec:result}
In this section, we present the results of our quantitative and qualitative studies, and answer our research questions.



\subsection*{RQ1: What are the most important factors associated with CI Build Failures at Atlassian?}
To investigate which factors influence CI build failures, we perform a quantitative analysis.
Table~\ref{tab:stat} provides the descriptive statistics of explanatory variables in the studied datasets.
Before building a model, we evaluate the correlation among explanatory variables based on hierarchical clustering analysis.
Then, we perform a redundancy analysis to detect and remove the redundancy.
We find that there are no explanatory variables that are highly correlated or redundant.
Hence, we use all explanatory variables to construct our models.

We build a Logistic Regression Model (LRM) to fit the explanatory variables.
Our model achieves an AUC of 0.82.
Moreover, the optimism of AUC is very small ($|Optimism| = 2e^{-5}$).
This result indicates that our model is stable and has a meaningful and robust explanatory power.
Table~\ref{tab:important} shows the explanatory power (Wald $\chi^2$ ) of our explanatory variables.
The $\chi^2$ statistic of each variable shows the proportion of the Wald $\chi^2$ of the entire model fit that is attributed to that explanatory variable.
The larger the proportion of the Wald $\chi^2$ is, the larger the explanatory power that a particular explanatory contributes explanatory power to the fit of the model.
The explanatory variables in the \textit{repository properties dimension} account for the largest proportion of the Wald $\chi^2$.
Previous studies~\cite{chen2020buildfast, hassan2017change, saidani2022improving} point out that in open-source projects the characteristics of the historical builds in the repository are important indicators for CI build failures.
Our findings corroborate those studies under the industrial scenario.

\begin{table}[t]
\centering
\caption{Descriptive statistics of the studied factors. Histograms are in a log scale for values exceeding one.}
\label{tab:stat}
\begin{adjustbox}{max width=\columnwidth}
\begin{tabular}{cccccl}
\hline
                         & 1st Qu. & Median & Mean & 3rd Qu. & Histograms \\ \hline
\multicolumn{5}{l}{\textit{PR Dimension}}         &           \\ \hline
changes\_num             & 1.0       & 3.0      & 9.5   & 8.0      &
 \raisebox{-0.2\totalheight}{\includegraphics[width=15ex, height=2.5ex]{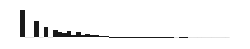}}     \\
comment\_num  & 0.0       & 1.0      & 4.0    & 4.0       &     \raisebox{-0.2\totalheight}{\includegraphics[width=18ex,height=2.5ex]{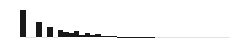}}       \\
reviewer\_num            & 2.0       & 4.0      & 5.1  & 7.0       &      \raisebox{-0.2\totalheight}{\includegraphics[width=16ex,height=2.5ex]{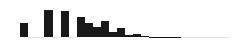}}      \\
wait\_time\_to\_review   & 0.0       & 1.0      & 18.1   & 13.0      &      \raisebox{-0.2\totalheight}{\includegraphics[width=14ex,height=2.5ex]{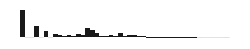}}      \\
reviewing\_time      & 0.0       & 1.0     & 38.9 & 22.0      &        \raisebox{-0.2\totalheight}{\includegraphics[width=14ex,height=2.5ex]{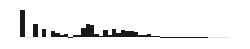}}    \\
per\_failed\_file & 0.0       & 0.2   & 0.4  & 0.8     &       \raisebox{-0.2\totalheight}{\includegraphics[width=12ex,height=2.5ex]{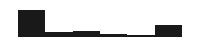}}     \\ \hline
\multicolumn{5}{l}{\textit{Repository Dimension}}     &          \\ \hline
repo\_prev\_build & \multicolumn{2}{c}{Failed: 29,316} & \multicolumn{2}{c}{Passed: 272,193} &  \\
repo\_rec\_build    & 0.0       & 0.2    & 0.2  & 0.2     &       \raisebox{-0.2\totalheight}{\includegraphics[width=12ex,height=2.5ex]{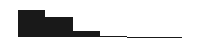}}     \\
repo\_hist\_build   & 0.0       & 0.0      & 0.1  & 0.2     &       \raisebox{-0.2\totalheight}{\includegraphics[width=12ex,height=2.5ex]{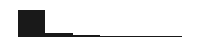}}     \\ \hline
\multicolumn{5}{l}{\textit{Contributor Dimension}}       &           \\ \hline
team\_num                & 1.0       & 1.0      & 1.5  & 2.0       &      \raisebox{-0.2\totalheight}{\includegraphics[width=20ex,height=2.5ex]{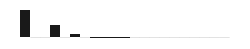}}      \\
team\_member\_num        & 7.0       & 10.0     & 15.3 & 15.0      &       \raisebox{-0.2\totalheight}{\includegraphics[width=16ex,height=2.5ex]{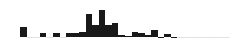}}     \\ \hline
\end{tabular}
\end{adjustbox}
\end{table}

\begin{table}[t]
\centering
\caption{Wald Statistics and partial effect of the logistic regression model for identifying CI build failures. The explanatory variables with higher percentage values of $\chi^{2}$ are more significant to the model. The larger the magnitude of the odds ratio is, the larger the partial effect that an explanatory variable has on the likelihood. The explanatory variables that contribute the most important explanatory power to a model (i.e., accounting for a large proportion of Wald $\chi^2$) are shown in boldface.}
\label{tab:important}
\begin{adjustbox}{max width=\columnwidth}
\begin{threeparttable}
\begin{tabular}{ccccc}
\hline
                         &  $\chi^2$     & $p$-value & Shifted Value & Odds Ratio \\ \hline
Wald  $\chi^2$                  & 45,235 & ***     & --            & --         \\ \hline
\multicolumn{5}{l}{\textit{PR Dimension}}                   \\ \hline
changes\_num             & 0.1\%    & ***     & 1.0→8.0           & 0.5\%↑       \\
comment\_num  & 0.1\%    & ***     & 0.0→4.0           & 1.6\%↑       \\
reviewer\_num            & 0.0\%    & **      & 2.0→7.0           & 2.2\%↓       \\
wait\_time\_to\_review   & 0.0\%    & ***     & 0.0→13.0          & 0.3\%↑        \\
reviewing\_time          & 0.0\%    & **      & 0.0→22.0          & 0.3\%↑        \\
per\_failed\_file & \textbf{1.4\%}    & ***     & 0.0→0.8        & 31.6\%↑      \\ \hline
\multicolumn{5}{l}{\textit{Repository Dimension}}                 \\ \hline
repo\_prev\_build  & \textbf{6.1\%}    & ***     & Passed→Failed & 174.7\%↑     \\
repo\_rec\_build    & \textbf{21.4\%}  & ***     & 0.0→0.2         & 111.0\%↑     \\
repo\_hist\_build   & \textbf{5.6\%}   & ***     & 0.0→0.2         & 29.6\%↑      \\ \hline
\multicolumn{5}{l}{\textit{Contributor Dimension}}                   \\ \hline
team\_num                & 0.2\%    & ***     & 1.0→2.0           & 5.8\%↑       \\
team\_member\_num        & 0.1\%    & ***     & 7.0→15.0          & 2.0\%↓       \\ \hline
\end{tabular}
\begin{tablenotes}
\item Statistical significance of explanatory power according to Wald $\chi^{2}$ likelihood ratio test: $^\circ p \geq 0.05; ^* 0.01 < p < 0.05; ^{**} 0.001 < p < 0.01; ^{***} p < 0.001$
\item ↑: The explanatory variable has a positive relationship with the response variable.
\item ↓: The explanatory variable has a negative relationship with the response variable.
\end{tablenotes}
\end{threeparttable}
\end{adjustbox}
\end{table}

\begin{figure}[t]
    \centering
    \includegraphics[width=\columnwidth]{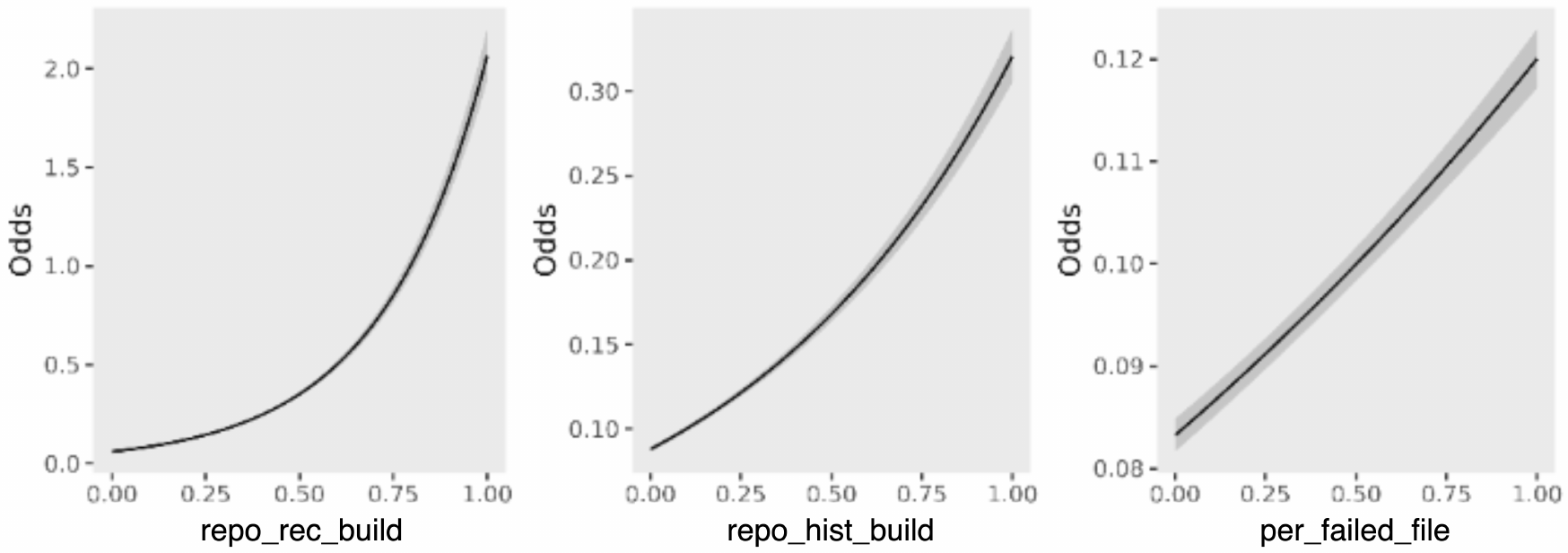}
    \caption{The relationship between the explanatory variable (x-axis) and the likelihood of a build failing (y-axis). The larger the odds value, the higher the likelihood that a build will fail. The gray area represents the 95\% confidence interval.}
    \label{fig:plots}
\end{figure}

To study the relationship between the explanatory variables and the response, we plot the odds produced by our models against each explanatory variable while holding the others at their median values.
Figure~\ref{fig:plots} shows the relationship between the explanatory variables and the response variable with the 95\% confidence interval (gray area).
To estimate the partial effect that the explanatory variables have on the likelihood, we analyze the relative change in the odds corresponding to a shift in the value of each explanatory variable.
Table~\ref{tab:important} shows the estimated partial effect of each explanatory variable in our models. 
The \textit{Odds Ratio} column shows the partial effect based on the \textit{shifted value}.
For continuous variables, the observed value is an interquartile range of those explanatory variables. For categorical variables, the observed value is a comparison between the observed categories.
Below, we present and discuss our analysis from the examination of these explanatory variables in relation to the response variable.

\textbf{\textit{The ratio of the builds that failed out of the recent five builds of the repository shares a positive relationship with the likelihood of CI build failure.}}
Table~\ref{tab:important} also shows that the ratio of the builds that failed out of the recent five builds in the repository shares a significant relationship with the likelihood that a build will fail.
The likelihood increases by 111\% when the ratio value increases from 0 to 0.2.
Figure~\ref{fig:plots} also shows an increasing trend in the likelihood of CI build failure as the value of this variable increases.
The narrowing of the confidence interval (gray area) suggests a high precision of this variable to determine the CI build failure.
This result indicates that a build tends to fail if many recent failures occur in the CI builds.

\textbf{\textit{The ratio of all historical builds in the repository that failed shares a positive relationship with the likelihood of CI build failure.}}
Figure~\ref{fig:plots} confirms this finding by illustrating an increasing trend between the likelihood and the value of the variable.
Table~\ref{tab:important} shows the likelihood increases by 29.6\% when the percentage value increases from 0 to 0.2.
This result indicates that a build tends to fail if many historical builds in the repository failed in the CI process.

\textbf{\textit{The likelihood of CI build failure increases if the previous build failed.}}
Table~\ref{tab:important} shows that the likelihood of the current CI build failing is 174.7\% higher in cases where the previous build fails, compared to where the previous build succeeds.
Our result indicates that if the previous build fails, the current build is likely to fail too.

\textbf{\textit{The ratio of changed files in the
PR that were involved in the past failed builds shares a positive relationship with the likelihood of CI build failure.}}
Table~\ref{tab:important} shows that the percentage of changed files in the current PR that were involved in the previous failed builds has a significant relationship with the likelihood of CI build failure.
The partial effect analysis shows that the likelihood can increase by 31.6\% when the value of the ratio increases from 0 to 0.8. 
Figure~\ref{fig:plots} also shows a positive trend between the likelihood of CI build failure and the value of this variable.
This result indicates that a build with a high ratio of files that previously failed before is likely to fail in the CI build.


\resultBox{\textbf{RQ1: The explanatory variables in the repository dimension share strong associations with the CI build outcomes. The ratio of the builds that failed out of the recent five builds of the repository, the ratio of all historical builds of the repository that failed, the status of the previous PR, and the ratio of changed files in the PR that were involved in the past failed builds share strong associations with CI build failures.}}

\subsection*{RQ2: What are practitioners’ perceptions about CI build failures?}

To answer RQ2, we asked three questions in our survey to capture insights from experienced developers who actively performed CI builds.
Q1 is a five-point Likert question that aims to quantify the level of challenges faced by developers when resolving the CI build failures.  
Q2 is an open-ended question to ask the impact of CI build failures.
Q3 is a multiple-selection question with optional free-text
responses to gather developers' opinions about the causes behind CI build failures occurring during the CI builds triggered on the main branch.

According to the results of Q1, 46\% of respondents perceived resolving CI build failures as very to extremely challenging (9\% extremely challenging, 37\% very challenging), whereas 13\% did not find it challenging. 
The remaining respondents rated the difficulty as moderate.
This result highlights that resolving CI build failures is challenging in software development at Atlassian.
The responses to Q2 revealed that CI build failures can impact both individual workflow and the overall organization.
Developers perceived that CI build failures may \textit{\textbf{increase time costs}} and even \textit{\textbf{disrupt the workflow}}.
As explained by D12, \textit{"When a CI build crashes, you have to play detective on what went wrong. It's a pain for you because it eats up time"}. 
D1 also pointed out that \textit{"CI build failures can slow down the development process and lead to delays in software releases"}.
Another impact of CI build failure is \textit{\textbf{reducing productivity}} of developers. 
This was because \textit{"someone needs to spend time looking at why it failed, fix it. Has potential to make roll backs or hot fixes take longer and affect time to release"} (D53).
D9 also explained, \textit{"It can really impact development productivity as it takes a while to find out if builds pass and then have to go back and make changes"}.
In addition, some developers believed CI build failures also have an impact on the \textit{\textbf{team}}.
D10 responded that CI build failure \textit{"can sometimes cause a bit of friction among teams, especially if we're working on tight deadlines"}.
Another response from D33 argued that CI build failures also highlight \textit{"[..] a need for better training and skill development within the team[..]"} because CI build failures \textit{"[..] might indicate knowledge gaps or a lack of familiarity with the project"} (D33).

\begin{figure*}[t]
    \centering
    \includegraphics[width=.8\textwidth]{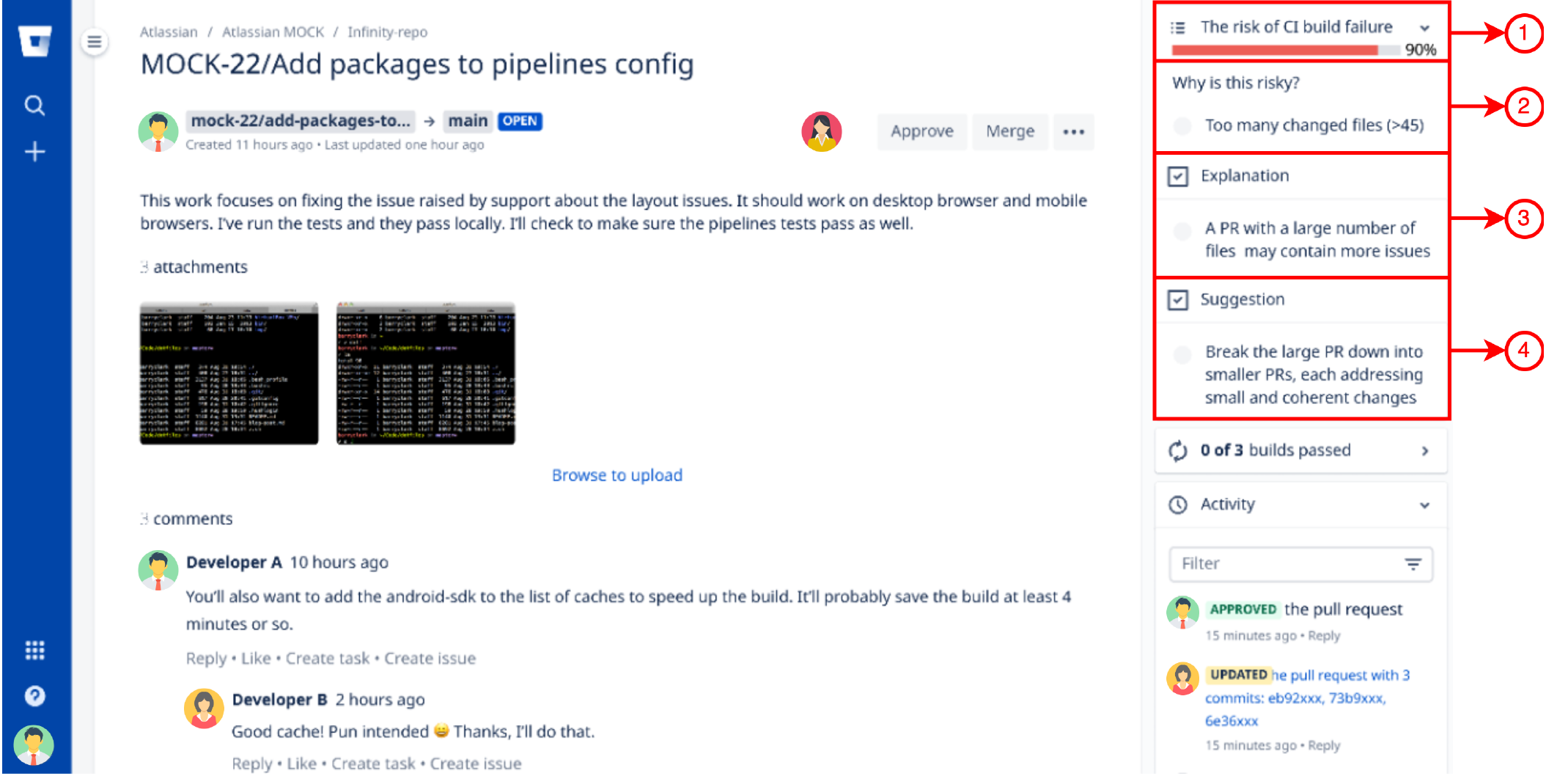}
    \caption{A UI prototype with mocked data and the prediction of the CI build outcome. (1) shows the likelihood that the build will fail. (2) suggests the factors that are associated with the CI build failure. (3) explains the factor that may result in CI build failure. (4) provides a suggestion that may reduce the likelihood of CI build failure.}
    \label{fig:mockup}
\end{figure*}

Despite the possible impact on the software development process of CI build failure, several developers believed that CI build failures are \textit{\textbf{learning opportunity}} for developers.
For example, D32 noted that  \textit{"CI build failures [..] are great learning chances to gain experience.."}.
Moreover, CI build failure is useful in \textit{\textbf{improving build processes}}.
By taking lessons from these CI build failures, developers could stay proactive: \textit{"[..] improve build processes. Staying proactive helps us avoid these issues in major patches later on [..]"} 
 (D32);  \textit{"These failures can highlight areas where processes need to be improved or updated"} (D34).


The result of Q3 shows that 77\% of respondents identified \textit{code issues} as the primary reason for CI build failures.
It suggests that developers perceived the improvement in code quality could aid in reducing CI build failures.
This underscores the necessity for development teams to invest in code quality assurance measurement such as the rigorous code review practices.
Furthermore, this finding is aligned with previous quantitative research~\cite{vassallo2017tale}, emphasizing the critical role of code quality in the CI build process.

\resultBox{\textbf{RQ2: Respondents perceived that CI build failures are challenging to resolve.
The responses from developers noted that CI build failures delay the software development process and reduce team productivity.
But they also provide opportunities for developers to take the lessons from failing and avoid later more serious issues.}}


\subsection*{RQ3: What are practitioners' perceptions on the usefulness of CI build prediction?}

To understand the perceived usefulness of the CI build prediction, Q4 aims to ask developers about the usefulness of this technique in their day-to-day workflow. 
To facilitate a clearer understanding, we developed a UI prototype (i.e., user interface) to visually demonstrate our concept.
Figure~\ref{fig:mockup} depicts our UI prototype that integrates the CI build prediction (red boxes)into Bitbucket.
The prediction consists of four main components: the likelihood, associated factors, an explanation, and a suggestion.

The responses to the Likert-scale question of Q4 show that 58\% of the respondents believed that the CI build prediction is somewhat to extremely useful (41\% extremely useful, 17\% somewhat useful).
Conversely, 29\% of the respondents argued that this technique may be not useful in practice.
The remaining respondents maintained a neutral stance.
This result indicates that CI build predictions could be useful in addressing CI build failures in practice.

The analysis of the answers to the open-ended question revealed that developers perceive the CI build prediction as providing \textit{\textbf{proactive insights}} into code quality control of PRs.
D12 regarded this technique as \textit{"a checkpoint"} for catching code errors in PRs. 
Developers appreciated its ability to \textit{"proactively identify potential issues"} (D19) and know \textit{"potential issues beforehand"} (D26).
With the help of this technique, developers could \textit{"avoid the bad code being merged to the codebase"} (D25) and \textit{"find the buggy code before the merge happens"} (D12).
Another highlighted benefit is its role in aiding \textit{\textbf{decision-making}}.
This aspect was further elaborated by D53, who commented on the technique's usefulness in \textit{"evaluating how teams can reduce risk and speed up development flow"}.
Such insights are especially beneficial in a fast-paced development environment where decision-making is important.

\begin{table*}[]
\caption{The factors, explanations, and examples of suggestions provided in the survey for the percentage of current changed files in previous failed builds (\textbf{\textit{per\_failed\_file}}), the number of changed files (\textbf{\textit{changes\_num}}), and the number of comments (\textbf{\textit{comment\_num}}).}
\label{tab:suggest}
\begin{adjustbox}{max width=0.8\textwidth}
\begin{tabular}{lll}
\hline
Factors & Explanations & Suggestions \\ \hline
per\_failed\_file &
  \begin{tabular}[c]{@{}l@{}}High percentage of changed files that were involved in the \\ previous failed build may cause the previous issues again.\end{tabular} &
  \begin{tabular}[c]{@{}l@{}}Ensure all files in previous failed build\\ are carefully checked.\end{tabular} \\ \hline
changes\_num &
  \begin{tabular}[c]{@{}l@{}}A PR with a large number of changed files may contain \\ more issues.\end{tabular} &
  \begin{tabular}[c]{@{}l@{}}Break the large PR down into smaller PRs, \\ each addressing small and coherent changes.\end{tabular} \\ \hline
comment\_num &
  \begin{tabular}[c]{@{}l@{}}A high number of comments may suggest more potential\\ issues or disagreements in the PR.\end{tabular} &
  Ensure all comments are addressed. \\ \hline
        &              &                       
\end{tabular}
\end{adjustbox}
\end{table*}

However, developers were concerned about the \textit{\textbf{accuracy}} of the predictions.
D27 raised a valid point about potential setbacks caused by \textit{"false positives and false negatives"}, which can lead to \textit{"delay our review and testing process".}
Another concern from D27 was the possibility of developers \textit{\textbf{relying too much on this technique}} \textit{"There is a risk of becoming overly reliant on these predictions"} if the predictions are seen as substitutes during code review practices.
D53 also pointed out that the utility of this technique may be \textit{\textbf{limited due to the different skill levels}} of developers, suggesting that it may be primarily beneficial to \textit{"junior members"} who may rely more on such techniques for guidance, as opposed to experienced developers who may find less value in it.
This distinction underscored the need for techniques that provide balanced and valuable insights for all members with different skill levels within a development team.

\resultBox{\textbf{RQ3: Respondents acknowledged the value of the CI build prediction in providing proactive insight into PRs and facilitating decision-making. However, the practitioners still had concerns about the accuracy, the risk of over-reliance, and providing balanced and valuable support to all developers across different skill levels.}}

\subsection*{RQ4: What are practitioners’ perceptions on the explanations and suggestions of CI build predictions?}

This RQ will focus on the practitioners' perceptions of explanations and suggestions based on the factors in the PR dimension. This is because these factors are within the scope where developers can take action on the currently submitted PRs.
Despite the strong associations of repository dimension shown in RQ1, these factors primarily measure the past builds which provides limited scope for developers to take any actions to improve the current PR to pass the build. 
Similarly, the factors in the contributor dimension are related to personnel, and the developer is unable to change them to reduce the likelihood of a CI build failure.  

We select three PR-dimension factors that have explanatory power greater than zero, i.e., the number of changed files, the number of comments, and the ratio of changed files in the PR that were involved in the past failed builds.
Then, we employ a Likert-scale question followed by an open-ended question to inquire about the perception of developers on these three factors (Q5-Q7). 
Table~\ref{tab:suggest} describes the \textit{explanation} and \textit{suggestion} corresponding to each factor presented in our survey.
We display the corresponding explanation and the suggestion for each factor in our UI prototype (see Fig~\ref{fig:mockup}). 


\begin{figure}[t]
    \centering
    \includegraphics[width=\columnwidth]{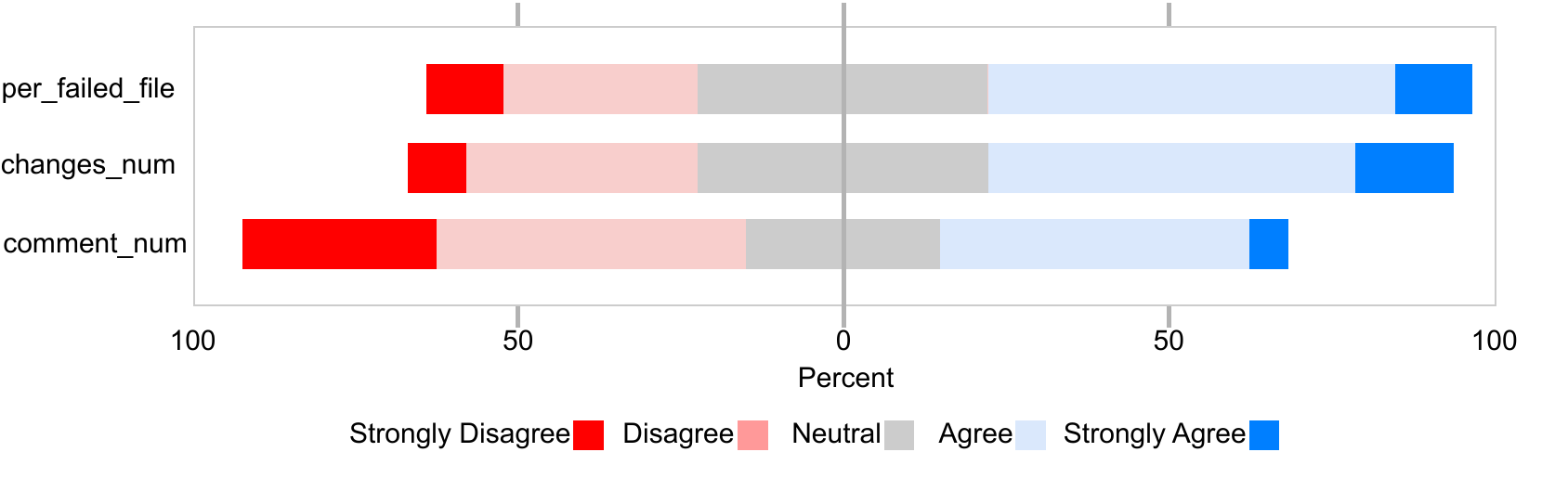}
    \caption{Practitioners' agreement on the explanations and suggestions based on PR-dimension factors.}
    \label{fig:factor}
\end{figure}

Fig~\ref{fig:factor} shows that 46\% of the respondents agreed with the explanation and suggestion based on the \textit{per\_failed\_file} factor, i.e., the  ratio of changed files in the PR that were involved in the past failed builds (7\% strongly agree, 39\% agree).
The respondents noted that the suggestion should  \textit{\textbf{specify the changed files in previous failed builds}}.
They expressed that listing these files in the suggestion could support: \textit{"make sure we're not missing any files that could trip us up later"} (D12) and \textit{"point out some potentially dangerous areas that should be focused on in review"} (D51).
Another important theme pointed out by the respondents is \textit{\textbf{identifying the specific issues in the changed files}}.
They showed a keen interest in understanding the root causes of the previous issues in these changed files that influence the CI build outcome. 
They believed that such insights could \textit{"help in making a more informed decision"} (D26), thereby enhancing the overall efficiency and success rate of CI builds.

Fig~\ref{fig:factor} shows that 44\% of respondents agreed with the explanation and suggestion based on the \textit{change\_num}, i.e., the number of changed files (9\% strongly agree, 35\% agree).
The open-ended question reveals that the respondents generally perceived \textit{big changes} in PRs as a \textit{risk} (D25) and \textit{a cause for concern} (D17) for CI build failure.
Some developers acknowledged that the provided suggestion in Table~\ref{tab:suggest} could reduce the CI build failure caused by a large number of changed files.
For instance, D42 noted that \textit{"Breaking big PR into small one is a good practice"}, while D19 added that \textit{"smaller updates are easier to handle and less likely to mess things up"}.
Nevertheless, some respondents noted that breaking down large PRs needs to be \textit{\textbf{evaluated on a case-by-case basis}}.
D27 elaborated on this and further stated that \textit{"sometimes, large updates are necessary and breaking them down isn't feasible"}.
This suggests scenarios where significant system modifications are required, and these changes need to be implemented together in a single large update.
This perspective emphasizes that explanations and suggestions of the factors depend on the context of the scenario.

Lastly, Fig~\ref{fig:factor} shows that the percentage of respondents who agree with the explanation and suggestion based on the impact of comment count is less than the prior explanations and suggestions (4\% strongly agree, 30\% agree). 
A key theme emerging from these responses centered around \textit{\textbf{the quality of comments}} rather than their quantity.
Many developers argued that it's the relevance and substance of comments rather than their number that might influence CI build failure.
For example, a high number of comments may be \textit{discussions} (D17) or \textit{conversation} (D25) that are not related to the CI build failure.
D40 shared a similar opinion that comments like "Nice" and "Great change", despite the frequency of these comments, do not contribute to understanding or preventing CI build failures. 

Contrary to our findings in quantitative analysis, some developers believed that \textit{\textbf{a high number of comments does not imply negative implications for the CI build}}.
D38 believed that \textit{"More comments does not mean more bad"}.
Some developers believed that a high number of comments may indicate \textit{"a thorough review"} (D27) and is\textit{"not an unhealthy process"} (D39).
From this perspective, a PR with a high number of comments could also give the \textit{"high confidence"} (D38) of the developers that the build will pass.  

\resultBox{\textbf{RQ4: Respondents agreed that CI build failures are associated with the percentage of current changed files that were involved in the previous failed builds and the number of changed files. 
However, while recognizing the significance of these factors in understanding the CI build failure, developers emphasized that the explanations and suggestions for these factors should be optimized to fit the context of development scenarios. 
This underscores a need for further research and investigation to refine and tailor these insights for varied scenarios.}}



%% file: sections/discussion.tex
\section{Implications} \label{sec:discussion}
Below, we provide the discussion about the implications of our work to researchers and practitioners.

\subsection{Suggestions for Researchers}
Beyond our research questions, we provide the following implications for researchers that deserve further study:

\textbf{Collaboration between Human Judgment and Automated Predictions}:
RQ3 shows the necessity of a balanced collaboration between human judgment and automated predictions of build outcomes in CI processes.
There's a need for detailed studies that explore the dynamics of trust, reliance, and decision-making between developers and CI build prediction.
For example, researchers can optimize the integration of these techniques by studying how developers interact with and interpret the predictions and suggestions provided.
Such research could involve developing frameworks that facilitate effective communication between the CI build prediction technique and the developer, ensuring that the predictions are not only accurate but also presented in a manner that is aligned with the developer’s cognitive and decision-making processes.

\textbf{Investigation into Contextual Dependencies}:
RQ4 highlights that the effectiveness of CI build predictions depends on the context, such as project characteristics and the individual experience levels of developers.  
This observation underscores the potential benefits of tailoring the level of automation and the nature of interactions between the build prediction technique and its users based on these contexts.
It suggests that a one-size-fits-all approach may not be optimal and that customization could enhance the build prediction's effectiveness and user satisfaction.
It is worth conducting further research involving in-depth case studies or comparative analyses across diverse contexts to understand the contextual factors affecting the utilization of CI builds prediction in practice.

\textbf{Enhancement of Build Prediction Explainability}:
Our observation from RQ4 pinpoints the importance of explainability in CI build prediction.
We found that developers are more likely to trust and effectively utilize these techniques when they understand the rationale behind predictions.
Although we give a glance into the explanation of the CI build prediction, there's substantial room for advancement.
Future research can focus on creating intuitive methods to convey complex model decisions such as interactive visualizations or clear narrative explanations.
Studies could also explore the impact of different explanation types on developer trust and model utilization, aiming to tailor explainability features to developers' specific needs and preferences.
Enhancing explainability plays an important role in fostering relationships between developers and CI build prediction.

\subsection{Implications for Practitioners}
Based on the findings and responses from the survey with Atlassian developers, we derive the following implications for practitioners.

\textbf{Encouraging Learning from CI Build Failures}:
RQ2 shows that although CI build failures can be disruptive, they are also seen as opportunities for learning from failures and improving build processes. 
Practitioners should draw lessons from these failures. 
This involves not just fixing the immediate issues but also understanding the underlying causes and taking steps to prevent similar issues in the future.

\textbf{Improvement of Code Review Practices}:
Aligning with the findings of prior research~\cite{rahman2017impact,zampetti2019study,cassee2020silent}, our study confirms that meaningful and constructive code review practices are essential for a successful CI process based on the findings of RQ2. 
Practitioners should encourage teams to engage in thorough code reviews, focusing on the substance of code changes. 
Investing time in comprehensive reviews can preempt issues that lead to build failures.

\textbf{Optimizing CI Processes with CI Build Prediction}:
The results of RQ3 indicate that developers appreciate the CI build prediction as a proactive measure for catching issues in builds.
It's recommended that organizations integrate such technique into their CI processes.
This technique not only identifies potential issues before code merging but also aids in decision-making, especially in fast-paced development environments where the cost of a delayed reaction can be considerable.

\textbf{Balancing Reliance with Human Judgment}:
From the results of RQ3, we found that while CI build prediction are valuable, there's a risk of developers becoming overly reliant on these techniques. 
It's recommended that while these techniques should be used to aid the review process, they should not replace the need for thorough human review. 
Practitioners should foster an environment where techniques are used as a complement to human expertise.

%% file: sections/threat.tex
\vspace{-1mm}
\section{Threats To Validity} \label{sec:threat}

\textit{Internal validity} concerns the quality of the study design and the rigorousness of its execution. 
In our study, these threats are related to data mining, model construction, as well as survey design and analysis. 
We directly utilized the aggregated data sourced from Atlassian's developer database (excluding any of Atlassian's customers' data). 
This provided us with access to the latest and most relevant dataset.
To enhance data accuracy, we conducted a manual inspection and filtered out the apparent outliers from the dataset.
The selection of factors for the LRM model may be a potential threat to our study. 
We mitigated this by relying on the metrics previously established and validated by the research community.
Other factors (e.g., human factors) may also play a role in CI build outcomes.
However, the selection of each factor is governed by the limitation of data availability.
Thus, we cannot analyze such factors due to limited access.
In designing our survey, we created questions to be both clear and easy to understand. 
While acknowledging some potential biases in analyzing open-ended questions, we addressed this by manually conducting a thematic analysis approach for unbiased analysis and engaging multiple coders to ensure an accurate interpretation of the data.
In addition, the developers facing significant challenges in CI builds may be more motivated to participate in surveys.
To address this bias, we invite a wide range of developers within Atlassian, not solely those who experienced CI build failures.

\textit{External validity} concerns the generalizability of our findings.
Our study specifically focuses on the projects developed by Atlassian teams and the developers involved in these projects. 
Although these projects are used in an industry setting and the survey participants are highly experienced full-time software developers, it's important to note that our findings may not be generalized across all other projects from other companies or those within open-source communities.
However, further research is essential to deepen the understanding of the CI build failures and to create a unified body of empirical knowledge.

%% file: sections/related.tex
\vspace{-2mm}
\section{Related Work} \label{sec:related_work}

In this section, we discuss the related work related to CI build failures and CI build predictions.

\textbf{CI Build Failures.}
CI build failures can disrupt the development work of entire teams for a considerable time~\cite{kerzazi2014automated,maipradit2023repeated}.
Such long build times greatly increase the cost of human and computing resources, and hence become a common barrier faced by software organizations adopting CI~\cite{hilton2017trade}.
To address CI build failures, many studies aim to understand the underlying causes~\cite{fu2024ai}.
Seo et al.~\cite{seo2014programmers} explored compiler errors that occur in the build process. 
They developed a classifier of compilation errors leading to build failures. 
Kerzazi et al.~\cite{kerzazi2014automated} examined the impact of socio-technical congruence on build failures.
They observed a high percentage of build failures (17.9\%) that brings a potential cost of about 2,035 man-hours, considering that each failure needs one hour of work to succeed.
Jin et al.~\cite{jin2022builds} developed a set of rules that indicates the safe builds to skip.
Similarly, Abdalkareem et al.~\cite{abdalkareem2019commits} manual investigate 588 java projects. 
They proposed the rules to identify the CI skipped commit.
Rausch et al.~\cite{rausch2017empirical} and Beller et al.~\cite{beller2017oops} focused on the impact of failed test cases on CI build. 
They highlight the importance of testing in CI build failure.
Orellana Cordero et al.~\cite{orellana2017differences} extended this focus to test-related build failures in open-source projects.
Gallaba et al.~\cite{gallaba2018noise} investigate 3.7 million build jobs to analyze the noise and heterogeneity in build outcome data.

Different from prior studies, our work focuses on investigating CI build failures within the context of Atlassian's development environment.
We explore both technical factors and the human and organizational factors that potentially influence CI build failures.
This enables us to uncover a broader spectrum of potential causes and impacts of CI build failures.

\textbf{CI Build Predictions.} Recent works leverage various techniques to predict the outcomes of CI builds. 
Xia and Li~\cite{xia2017could} explored 6 predictive models across 126 open-source projects to predict the CI build outcome.
Finlay et al.~\cite{finlay2014data} used data stream mining techniques based on code metrics (i.e., basic metrics, dependency metrics, complexity metrics, cohesion metrics, and Halstead metrics) to predict build outcomes.
Jin et al.~\cite{jin2020cost} proposed \textsc{SmartBuildSkip} to predict the first build in a sequence of build failures.
Abdalkareem et al.~\cite{abdalkareem2020machine} proposed a machine learning approach to determine the commits that can be CI skipped.
Barrak et al.~\cite{barrak2021builds} performed a replication study using different datasets and features to explain and predict the outcome of builds.
Jin and Servant~\cite{jin2023hybridcisave} introduce a hybrid technique called \textsc{HybridCISave} to predict CI build failure.
Ni and Li~\cite{ni2017cost} predicted build outcomes in CI based on file-level metrics from the current and previous builds and failure statistics from historical builds. 
Hassan and Wang~\cite{hassan2017change} leveraged metrics from the current and previous build.
Chen et al.~\cite{chen2020buildfast} also extract the metrics from historical builds for their prediction.
Our findings further emphasize the vital role that factors related to historical builds have in predicting CI build failures.
This confirms that these prior findings derived from open-source projects are applicable to the industrial context of Atlassian.
In addition, some researchers~\cite{wolf2009predicting,schroter2010predicting} integrated social network analysis to capture communication structure.
They incorporate these structures into models to predict build outcomes. 

Despite recent advances, build outcome prediction still faces 
challenges that limit the practical application of these techniques in CI.
Firstly, the explanation of the predictive model needs to be investigated.
The effectiveness of the predictive model hinges on their ability to provide insightful explanations~\cite{tantithamthavorn2021explainable,jiarpakdee2020empirical,jiarpakdee2021practitioners,tantithamthavorn2015impact, pornprasit2021pyexplainer, tantithamthavorn2020explainable, tantithamthavorn2021actionable, tantithamthavorn2023explainable}.
To help developers effectively address CI build failures, it is necessary to explain the models of these CI build predictions and understand the reasoning behind the predictions.
Secondly, there's a gap in understanding how developers perceive and interact with the CI build prediction techniques.
Grasping developers' expectations and concerns is crucial for the successful integration and adoption of these techniques in real-world scenarios.
Thus, we recognize our work as a complement to those previous works.

%% file: sections/conclusion.tex
\vspace{-2mm}
\section{Conclusion}\label{sec:conclusion}
In this work, we investigated the CI build failures within Atlassian's projects, highlighting the challenges of addressing CI build failures in a real-world software development environment.
Our analysis revealed that the factors in the repository dimension are strongly associated with the CI build outcomes.
In addition, Atlassian's developers perceived that CI build failures are challenging to resolve.
They believed that CI build failures could delay the software development process and reduce team productivity. 
Moreover, we identified that both technical and human factors play significant roles in adopting the CI build prediction in practice. 
We hope that the findings and the insights we revealed can improve CI build failure prediction and aid developers in performing the CI build.

\section{Disclaimer} The perspectives and conclusions presented in this document are solely the authors' and should not be interpreted as representing the official policies or endorsements of Atlassian or any of its subsidiaries and affiliates. Additionally, the outcomes of this study are independent of, and should not be construed as an assessment of, the quality of products offered by Atlassian.
